\documentclass[floatfix,showpacs,showkeys,preprint,aps,showpacs,amsmath,amssymb,10pt]{revtex4-1}
\usepackage{cleveref}
\usepackage{multirow}
\usepackage{mathtools}
\usepackage{subfigure}
\usepackage[utf8]{inputenc}
\usepackage{color}
\usepackage{xcolor}
\usepackage{float}
\usepackage{braket}
\usepackage{dcolumn}
\usepackage{bm}
\usepackage{ulem}
\usepackage{cancel}
\usepackage{graphicx,epsf,epsfig}
\usepackage{footnote}
\usepackage{subfigure}
\usepackage{lipsum}
\usepackage{multirow}
\usepackage{tabulary}
\usepackage{rotating}
\usepackage{array}
\usepackage{slashed}
\usepackage[usenames, dvipsnames]{pstricks}
\usepackage{footnote} 

\definecolor{greenLinks}{rgb}{0,0.6,0}
\definecolor{blueLinks}{rgb}{0,0,0.6}
\definecolor{redLinks}{rgb}{0.6,0,0}
\definecolor{tempText}{rgb}{0.55,0.10,0.67}
\definecolor{eprintLinks}{rgb}{0.4,0.4,0.4}
\definecolor{journalLinks}{rgb}{0.6,0,0}

\usepackage{soul}

\begin{document}

\title{Single Higgs boson production in association with a top quark through FCNSI}
	
\author{V. M. L\'opez-Guerrero}
\email{victor.lopezguer@alumno.buap.mx}
\affiliation{\small Facultad de Ciencias F\'isico-Matem\'aticas, Benem\'erita Universidad Aut\'onoma de Puebla, C.P. 72570, Puebla, M\'exico,}
\affiliation{Centro Internacional de F\'isica Fundamental (CIFFU), Benem\'erita Universidad Aut\'onoma de Puebla, C.P. 72570, Puebla, M\'exico.}
	
\author{M. A. Arroyo-Ure\~na}
\email{marco.arroyo@fcfm.buap.mx}
\affiliation{Facultad de Ciencias F\'isico-Matem\'aticas, Benem\'erita Universidad Aut\'onoma de Puebla, C.P. 72570, Puebla, M\'exico,}
\affiliation{Centro Interdisciplinario de Investigaci\'on y Ense\~nanza de la Ciencia (CIIEC), Benem\'erita Universidad Aut\'onoma de Puebla, C.P. 72570, Puebla, M\'exico.}
\author{J. L. D\'iaz-Cruz }
\email{jldiaz@fcfm.buap.mx }
\affiliation{\small Facultad de Ciencias F\'isico-Matem\'aticas, Benem\'erita Universidad Aut\'onoma de Puebla, C.P. 72570, Puebla, M\'exico,}
\affiliation{Centro Interdisciplinario de Investigaci\'on y Ense\~nanza de la Ciencia (CIIEC), Benem\'erita Universidad Aut\'onoma de Puebla, C.P. 72570, Puebla, M\'exico.}
\author{O. F\'elix-Beltr\'an }
\email{olga.felix@correo.buap.mx }
\affiliation{Facultad de Ciencias de la Electr\'onica}
\affiliation{Centro Internacional de F\'isica Fundamental (CIFFU), Benem\'erita Universidad Aut\'onoma de Puebla, C.P. 72570, Puebla, M\'exico.}
\author{T. A. Valencia-P\'erez}
\email{tvalencia@fisica.unam.mx}
\affiliation{Instituto de F\'isica, 
Universidad Nacional Aut\'onoma de M\'exico, C.P. 01000, CDMX, M\'exico.}

\begin{abstract}
We study the production and possible detection of a single Higgs boson in association with a top quark in proton-proton collisions ($pp \to th + X$) at the High-Luminosity Large Hadron Collider. This process absent in the Standard Model is predicted by other models such as the Two-Higgs Doublet Model of type III, which is the theoretical framework adopted in this work. Promising results are found for specific scenarios of the model parameter space, which consist mainly of the parameters $\tan\beta$, $\cos(\alpha-\beta)$ and the parameter $\chi_{tc}$, responsible for the Flavor-Changing Neutral Scalar Interactions (FCNSI). Using the machine learning \textit{Boosted Decision Trees} algorithm and considering a systematic uncertainty of $5\%$, we predict \textit{signal significances} at level of $5\sigma$ for $\tan\beta=1$, $\cos(\alpha-\beta)=0.1$, $\chi_{tc}=5$, and integrated luminosities {$\mathcal{L}_{\rm int}\gtrsim2700~fb^{-1}$}. Likewise, we also predict a \textit{signal significance} $\sigma\approx 4.4$ for $\tan\beta=1$, $\cos(\alpha-\beta)=0.1$, $\chi_{tc}=3$, and integrated luminosities $\mathcal{L}_{\rm int} = 3000~fb^{-1}$, which consider the upper limit given by HL-LHC projection  on  $\mathcal{BR}(t\to ch)$.
\end{abstract}

\maketitle

\section{Introduction\label{sec:introduction}}

More than twelve years ago, the ATLAS and CMS collaborations announced the observation of a new spin-0 particle with mass $m=125.10 \pm 0.14\,\textrm{GeV}$ consistent with the Higgs boson properties predicted by the Standard Model (SM)~\cite{{Aad:2012tfa},{Chatrchyan:2013lba}}. After the discovery of the Higgs boson, an objective of the Large Hadron Collider (LHC) is to characterize its properties because, among other things, the Higgs field is fundamental to the generation of the particle masses contained in the SM~\cite{{PhysRevLett.13.508},{PhysRevLett.13.321}}. As we know, the SM provides a detailed description of the weak, strong and electromagnetic interactions, also successfully explains many of the experimental observations made in particle physics. However, despite these achievements, there are phenomena that cannot be explained within the SM, which is commonly referred as Beyond Standard Model (BSM) physics~\cite{Virdee:2016mzw, Lee2019, Miller, Boubaa2020, Crivellin:2023zui}. The fact that the SM leaves open questions for some phenomena directs our attention to exploring additional models, particularly those with an extended Higgs sector. One of them is the Two-Higgs Doublet Model of type III (2HDM-III)~\cite{Atwood:1997,Felix-Beltran:2013tra,Ghosh:2021jeg, Arroyo-Urena:2013cyf}, which predicts Flavor-Changing Neutral Currents (FCNC) at tree-level. The 2HDM-III leads to study interactions not found in the SM and to perform analysis of processes that could enhance the SM-like Higgs boson signals. It is also attractive to investigate Flavor-Violating (FV) processes involving top quarks because these kinds of interactions have large rates predicted by the model and at the same time allowed by experimental constraints. In contrast, FV processes are strongly suppressed in the SM due to the Glashow-Iliopoulos-Maiani (GIM) mechanism~\cite{Glashow:1970gm} and FCNSI are forbidden. Recently, the ATLAS~\cite{ATLAS:2022gzn} and CMS~\cite{CMS:2024ubt} collaborations reported upper limits on the $\mathcal{BR}(t\to ch)$, particularly CMS reported a $\mathcal{BR}(t\to ch)<0.00046$ at 95\% of confidence level, whose importance for this work lies in the fact that this result imposes limits on the parameter $\chi_{tc}$, responsible for the Flavor-Changing interaction in the process to study. The associated production $th$ for probing FCNC is proposed in Ref.~\cite{Aguilar-Saavedra:2000xbc}, a review of the process $pp \to th+X$ is presented in Ref.~\cite{Castro:2022qkg} while the authors of Ref.~\cite{Liu:2016gsi} studied the process from an effective field theory viewpoint, analyzing the impact of the coupling $\lambda_{tqh}$ on the production cross-section and the \textit{signal significance}. However, as shown later, machine learning techniques allow one to explore smaller values for the $tqh$ coupling ($g_{tqh}$).

In this paper, we are interested in the single Higgs boson production in association with a top quark via proton-proton collisions at the High Luminosity LHC (HL-LHC)~\cite{Apollinari:2015wtw}. Specific channels of decay are considered, that is,  $pp\to th+X\,(t\to \ell\nu_{\ell}b,\,h\to\gamma\gamma)$. Although the branching ratio of the Higgs boson into photons is relatively small $\mathcal{O}(10^{-3})$, this channel provides a good resolution on the Higgs mass and small QCD backgrounds, and is therefore essential for signal isolation, as will be seen later.

This work is structured as follows. In Sec.~\ref{sec:SecII}, we conduct a comprehensive review of the 2HDM-III with a particular emphasis on the theoretical implications of employing a four-zero texture in the Yukawa Lagrangian.  Experimental constraints on the model parameter space are also included. Section~\ref{sec:SecIV} focuses on taking advantage of the insights gained from previous sections, performing a computational analysis of the proposed signal and its SM background processes. Finally, the conclusions are presented in Sec.~\ref{sec:SecV}.

\section{The model \label{sec:SecII}}

The 2HDM introduces an additional Higgs doublet with the same hypercharge +1: $\Phi_a^T=( \phi_{a}^{+},\phi_{a}^{0})$ $(a=1, 2$), where $\phi^0_a$ denotes the neutral part and $\phi^\pm_a$ denotes the charged part. Depending of the vacuum expectation values (VEVs) chosen for the two Higgs doublets we can have two basis: ($i$) in the Higgs basis only one of the two Higgs doublets acquires a VEV, ($ii$) whereas in the generic basis both of the Higgs doublets acquire a VEV. The 2HDM built on the generic basis is called ``the general 2HDM''~\cite{Branco:2011iw}. In this section, we present the scalar potential and the Yukawa Lagrangian for the 2HDM type III.

\subsection{Scalar potential\label{subsec:scalarpotential}}

The most general $SU(2)_L \times U(1)_Y$ invariant scalar potential of the 2HDM is given by~\cite{Branco:2011iw,LorenzoDiaz-Cruz:2019imm}
\begin{equation}
\begin{array}{rcl}
V &=& \mu_{1}^{2}(\Phi_{1}^{\dag}\Phi_{1}^{}) + \mu_{2}^{2}(\Phi_{2}^{\dag}\Phi_{2}^{}) - (\mu_{12}^{2}(\Phi_{1}^{\dag}\Phi_{2}^{} + \textrm{H.c.})) \\ 
&+& \dfrac{\lambda_1}{2}(\Phi_{1}^{\dag}\Phi_{1})^2 + \dfrac{\lambda_2}{2}(\Phi_{2}^{\dag}\Phi_{2}^{})^2 + \lambda_{3}(\Phi_{1}^{\dag}\Phi_{1}^{})(\Phi_{2}^{\dag}\Phi_{2}^{}) \\
&+& \lambda_{4}(\Phi_{1}^{\dag}\Phi_{2}^{})(\Phi_{2}^{\dag}\Phi_{1}^{}) 
+ \Bigg( \dfrac{\lambda_5}{2}(\Phi_{1}^{\dag}\Phi_{2}^{})^2 \\
&+& (\lambda_{6}(\Phi_{1}^{\dag}\Phi_{1}^{}) + \lambda_{7}(\Phi_{2}^{\dag}\Phi_{2}^{}))(\Phi_{1}^{\dag}\Phi_{2}^{}) + \textrm{H.c.} \Bigg).
\label{eq:LPotential}
\end{array}
\end{equation}
Hermiticity condition for the potential implies that the $\mu_{1,2},$ $\lambda_{1,2,3,4}$ parameters are real, while the $\lambda_{5,6,7}$ and $\mu_{12}^2$ parameters can be complex numbers. We choose to work in a general 2HDM CP-conserving with a softly broken $Z_2$ symmetry.

After the Electroweak Symmetry Breaking (EWSB), the two Higgs doublets acquire VEVs given as
\begin{equation}
\braket{\Phi_a}=\frac{1}{\sqrt{2}}\left(\begin{array}{c}
0\\
\upsilon_a
\end{array}\right), \quad a= 1,2,
\end{equation}
where $v_{1,2}$ are real and satisfy $v=\sqrt{v^2_1 + v^2_2}=246$ GeV~\cite{ParticleDataGroup:2020ssz}. The VEV ratio defines the parameter $\tan\beta\equiv v_2/v_1$.
		
The mass matrix in the CP-even sector is given by
\begin{equation}
\mathcal{M}_{even} = 
\begin{pmatrix}
m_{11} & m_{12}\\
m_{12} & m_{22}
\end{pmatrix},
\label{eq:Meven}
\end{equation}
where
\begin{equation}
\begin{array}{rcl}
m_{11}&=&2\lambda_1v^2 \cos^2\beta+2\mu^2_{12}\dfrac{\sin\beta}{\cos\beta},\\
m_{12}&=&2(\lambda_3 +\lambda_4 +\lambda_5)v^2\cos\beta\sin\beta-\mu^2_{12},\\
m_{22}&=&2\lambda_2v^2 \sin^2\beta+2\mu^2_{12}\dfrac{\cos\beta}{\sin\beta}.
\end{array}
\end{equation}
After diagonalization of the mass matrix $\mathcal{M}_{even}$ and defining a mixing angle $\alpha$, we translate the gauge states into the mass eigenstates $h^0$ and $H^0$:
\begin{equation}
\begin{pmatrix} H^{0}\\
h^{0} 
\end{pmatrix} = 
\begin{pmatrix}
\cos\alpha & \sin\alpha \\
-\sin\alpha & \cos\alpha
\end{pmatrix}
\begin{pmatrix}
\operatorname{Re}(\phi^0_1)\\
\operatorname{Re}(\phi^0_2)
\end{pmatrix},
\label{eq:mixingangle}
\end{equation}
with
\begin{equation}
\tan2\alpha=\frac{2m_{12}}{m_{11}-m_{22}}.
\end{equation}	
Likewise, the mixing angle $\beta$ rotates the charged components of $\Phi_a$ into the charged mass eigenstates $G^\pm_W$ and $H^\pm$ as follows:
\begin{equation}
\begin{pmatrix} G^\pm_W\\
H^\pm
\end{pmatrix} = 
\begin{pmatrix}
\cos\beta & \sin\beta \\
-\sin\beta & \cos\beta
\end{pmatrix}
\begin{pmatrix}
\phi^\pm_1\\
\phi^\pm_2
\end{pmatrix}.
\label{eq:WHp_eigenstates}
\end{equation}
On the other hand, to obtain the neutral mass eigenstates $G_Z$ and $A$ we make the following rotation:
\begin{equation}
\begin{pmatrix} G_{Z^0}\\
A^0 
\end{pmatrix} = 
\begin{pmatrix}
\cos\beta & \sin\beta \\
-\sin\beta & \cos\beta
\end{pmatrix}
\begin{pmatrix}
\operatorname{Im}(\phi^0_1)\\
\operatorname{Im}(\phi^0_2)
\end{pmatrix}.
\label{eq:ZA_eigenstates}
\end{equation}	
We obtain the physical masses $M_{H^\pm}$, $M_{h^0}$, $M_{H^0}$ and $M_{A^0}$, given by the equations:
\begin{eqnarray}
M_{H^0,h^0}^2&=&\frac{1}{2}\Big(m_{11}+m_{22}\pm\sqrt{(m_{11}-m_{22})^2+4m_{12}^2}\Big),  
\\
M_{H^{\pm}}^2&=&\frac{\mu^2_{12}}{\sin\beta\cos\beta}-\frac{1}{2}v^2\Big(\lambda_4+\lambda_{5}\Big),
\\
M_{A^0}^2&=&M^2_{H^\pm}+\frac{1}{2}v^2(\lambda_4-\lambda_{5}).
\end{eqnarray}	
	
Four new particles emerge from the introduction of a second Higgs doublet, $H^0,\,A^0$ and $H^\pm$. Detection of these new particles is a goal of future experiments such as HL-LHC~\cite{Ashry:2023fmz} and FCC~\cite{Kuday:2017vsh,Das:2018dse}.

\subsection{Yukawa Lagrangian\label{subsec:yukawalag}}

The different versions of the 2HDM can be defined through the Higgs doublets and fermionic field interactions, as shown in Table~\ref{tab:2hdm_models}.

\begin{table}[H]
\centering
\begin{tabular}{|c|c|c|c|}
\hline
 Model & $u^i_R$&$d^i_R$&$l^i_R$\\ \hline
\multicolumn{4}{|c|}{Flavor conserving models}\\ \hline
I&$\Phi_2$&$\Phi_2$&$\Phi_2$\\
II&$\Phi_2$&$\Phi_1$&$\Phi_1$\\
Lepton-specific&$\Phi_2$&$\Phi_2$&$\Phi_1$\\
Flipped&$\Phi_2$&$\Phi_1$&$\Phi_2$\\ \hline
\multicolumn{4}{|c|}{Flavor violating model}\\ \hline
III&$\Phi_1$,$\Phi_2$&$\Phi_1$,$\Phi_2$&$\Phi_1$,$\Phi_2$\\
\hline
\end{tabular}
\caption{2HDM definitions based on the  Higgs doublets couplings with fermions in the Yukawa sector.}
\label{tab:2hdm_models}
\end{table}		

For the 2HDM-III, the Yukawa Lagrangian is written as \cite{PhysRevD.69.095002}:
\begin{equation}
\begin{array}{rcl}
-\mathcal{L}_Y &=&\Big( Y_{1}^{u} \overline{Q}_{L} \widetilde{\Phi}_{1} u_{R} + Y_{2}^{u} \overline{Q}_{L} \widetilde{\Phi}_{2} u_{R} + Y_{1}^{d} \overline{Q}_{L} \Phi_{1} d_{R} \\
&+& Y_{2}^{d} \overline{Q}_{L} \Phi_{2} d_{R}
 + Y_{1}^{l} \overline{L}_{L} \Phi_1 l_{R} + Y_{2}^{l} \overline{L}_{L} \Phi_2 l_{R} \\
 &+&\text{H.c.} \Big),
\end{array}
\label{eq:LYukawa} 
\end{equation}
where $\widetilde{\Phi}_{a} = i\sigma_2 \Phi_{a}^{*}$, $Q_L$ is the weak isospin quark doublet, $L_L$ is the weak isospin lepton doublet, $u_R$, $d_R$ are weak isospin quark singlets, $l_R$ is the weak isospin lepton singlet and $Y^f_a$ ($f=u,d,\ell$) are the $3\times 3$ Yukawa matrices.  

From the Yukawa Lagrangian in Eq.~\eqref{eq:LYukawa} and after the EWSB, the fermion mass matrices come into:
	\begin{equation}
		M_f = \frac{1}{\sqrt{2}} \left( v_1 Y_{1}^{f} + v_2 Y_{2}^{f} \right),\; f=u,d,\ell.
	\label{eq:Mfmatrix} 	
	\end{equation}
We assume that both Yukawa matrices have a four-zero texture~\cite{Diaz_Cruz_2005} and that they are Hermitian. The sum of such matrices inherits its structure to the mass matrix as follows
\begin{equation}\label{4zero_MASS_matrices}
M_f=\left(\begin{array}{ccc}0 & D_f & 0 \\D_f & C_f & B_f \\0 & B_f & A_f\end{array}\right),\;
\end{equation}
The elements of the matrix in Eq.~\eqref{4zero_MASS_matrices}
are related to the fermion masses $m_{f_i}$ ($i=1,\,2,\,3$), through the principal invariants:
\begin{equation}
\begin{array}{rcl}
\text{Tr}\left(M_f\right) & = & C_f+A_f=m_{f_1}+m_{f_2}+m_{f_3},\\
\lambda\left(M_f\right) & = & C_fA_f-D_f^{2}-B_f \\&=&m_{f_1}m_{f_2}+m_{f_1}m_{f_3}+m_{f_2}m_{f_3},\\ \text{det}\left(M_f\right) & = & -D_f^{2}A_f=m_{f_1}m_{f_2}m_{f_3}.
\label{Invariantes} 
\end{array}
\end{equation}
From Eq.~\eqref{Invariantes} we find a relation between the components of the mass matrix of four-zero textures and the fermion masses as follows:
\begin{eqnarray}\label{ElementosMatrizMasa}
A_f & = & m_{f_3}-m_{f_2},\nonumber \\
B_f & = & m_{f_3}\sqrt{\frac{r_{2}(r_{2}+r_{1}-1)(r_{2}+r_{2}-1)}{1-r_{2}},}\\
C_f & = & m_{f_3}(r_{2}+r_{1}+r_{2}),\nonumber \\
D_f & = & \sqrt{\frac{m_{f_1}m_{f_2}}{1-r_{2}}},\nonumber 
\end{eqnarray}
where $r_i=m_{f_i}/m_{f_3}$.
Performing a bi-unitary transformation we diagonalize the fermion mass matrices as
\begin{equation}\label{eq:MfmatrixDiag}
\overline{M}_f = V_{fL}^{\dag} M_f V_{fR}= \tfrac{1}{\sqrt{2}} \left( v_1 \widetilde{Y}_{1}^{f} + v_2 \widetilde{Y}_{2}^{f} \right),
\end{equation}
where $\overline{M}_f = \operatorname{diag}(m_{f_1},m_{f_2},m_{f_3})$ and $\widetilde{Y}_{a}^{f} = V_{fL}^{\dag} Y_{a}^{f} V_{fR}$. However, the Yukawa matrices are not necessarily diagonalized. As a consequence, FCNSI will be induced at tree-level.
The fact that $M_f$ is assumed as Hermitian, implies that $V_{fL}=V_{fR}\equiv V_f$. Here, $V_f=\mathcal{O}_{f}P_f$, where $P_f=\text{diag}\{e^{i\alpha_f},\,e^{i\beta_f},\,1\}$ and 
\begin{widetext}
\begin{equation}
\mathcal{O}_{f}=\left(\begin{array}{ccc}
\sqrt{\dfrac{m_{f_2}m_{f_3}(A-m_{f_1})}{A(m_{f_2}-m_{f_1})(m_{f_3}-m_{f_1})}} & \sqrt{\dfrac{m_{f_1}m_{f_3}(m_{f_2}-A)}{A(m_{f_2}-m_{f_1})(m_{f_3}-m_{f_2})}} & \sqrt{\dfrac{m_{f_1}m_{f_3}(A-m_{f_3})}{A(m_{f_3}-m_{f_1})(m_{f_3}-m_{f_2})}}\\
-\sqrt{\dfrac{m_{f_1}(m_{f_1}-A)}{(m_{f_2}-m_{f_1})(m_{f_3}-m_{f_1})}} & \sqrt{\dfrac{m_{f_2}(A-m_{f_2})}{(m_{f_2}-m_{f_1})(m_{f_3}-m_{f_2})}} & \sqrt{\dfrac{m_{f_3}(m_{f_2}-A)}{(m_{f_2}-m_{f_1})(m_{f_3}-m_{f_2})}}\\
\sqrt{\dfrac{m_{f_1}(A-m_{f_2})(A-m_{f_3})}{A(m_{f_2}-m_{f_1})(m_{f_3}-m_{f_1})}} & -\sqrt{\dfrac{m_{f_2}(A-m_{f_1})(m_{f_3}-A)}{A(m_{f_2}-m_{f_1})(m_{f_3}-m_{f_2})}} & \sqrt{\dfrac{m_{f_3}(A-m_{f_1})(A-m_{f_2})}{A(m_{f_3}-m_{f_1})(m_{f_3}-m_{f_2})}}
\end{array}\right).
\end{equation}
\end{widetext}
An important point is that $V_{u,d}$ must reproduce the observed matrix elements $V_{\text{CKM}}$, which are obtained as $V_{\text{CKM}}=V_{u}^{\dagger}V_{d}$, as shown in Ref.~\cite{Arroyo-Urena:2013cyf}. From Eq.~\eqref{eq:MfmatrixDiag}, we obtain the following expression
\begin{equation}
\big[ \widetilde{Y}_a^f \big]_{ij} = \frac{\sqrt{2}}{v_a}\delta_{ij}\overline{M}_{ij}^f-\frac{v_b}{v_a}\big[ \widetilde{Y}_b^f \big]_{ij}, \; b=1,2,
\label{eq:Yukawa_matrices}
\end{equation}
where
\begin{equation}
\big[ \widetilde{Y}_a^{f} \big]_{ij}
= \frac{\sqrt{m_{f_i} m_{f_j}}}{v} \, \big[\widetilde{\chi}_{a}^f \big]_{ij}.
\label{eq:cheng-sher}
\end{equation}
The $\widetilde{\chi}_{a}^f$ matrices are responsible for inducing Flavor-Changing (FC) interactions at three-level. Finally using Eqs.~\eqref{eq:LYukawa}-\eqref{eq:cheng-sher} and by making use of the definitions in Ref.~\cite{Hernandez-Sanchez:2012vxa}, we derive the Yukawa Lagrangian for quarks and the neutral scalar bosons $\mathcal{L}^Y_{ns}$ given by
\begin{align}
\label{eq:LYukawa_ns}
\mathcal{L}^Y_{ns}&=-\dfrac{g}{2 M_{W}}\\ 
& \times\Bigg[\bar{d}_{i}\Bigg(\left[m_{d_{i}} \xi_{H}^{d} \delta_{i j}-\dfrac{\left(\xi_{h}^{d}+X \xi_{H}^{d}\right)}{f(X)} \dfrac{\sqrt{m_{d_{i}} m_{d_{j}}}}{\sqrt{2}} \widetilde{\chi}_{i j}^{d}\right] H \nonumber\\
&+ \left[m_{d_{i}} \xi_{h}^{d} \delta_{i j}+\dfrac{\left(\xi_{H}^{d}-X \xi_{h}^{d}\right)}{f(X)} \dfrac{\sqrt{m_{d_{i}} m_{d_{j}}}}{\sqrt{2}} \widetilde{\chi}_{i j}\right] h \nonumber \\ 
&- i\left[m_{d_{i}} X \delta_{i j}-f(X) \dfrac{\sqrt{m_{d_{i}} m_{d_{j}}}}{\sqrt{2}} \widetilde{\chi}_{i j}^{d}\right] \gamma^{5} A \Bigg) d_{j} \nonumber\\
&+ u_{i}\Bigg(\left[m_{u_{i}} \xi_{H}^{u} \delta_{i j}+\dfrac{\left(\xi_{h}^{u}-Y \xi_{H}^{u}\right)}{f(Y)} \dfrac{\sqrt{m_{u_{i}} m_{u_{j}}}}{\sqrt{2}} \widetilde{\chi}_{i j}^{u}\right]  H \nonumber\\ 
&+\left[m_{u_{i}} \xi_{h}^{u} \delta_{i j}-\dfrac{\left(\xi_{H}^{u}+Y \xi_{h}^{u}\right)}{f(Y)}\left(\dfrac{\sqrt{m_{u_{i}} m_{u_{j}}}}{\sqrt{2}}\right) \widetilde{\chi}_{i j}^{u}\right] h \nonumber\\
&-i\left[m_{u_{i}} Y \delta_{i j}-f(Y) \dfrac{\sqrt{m_{u_{i}} m_{u_{j}}}}{\sqrt{2}} \widetilde{\chi}_{i j}^{u}\right] \gamma^{5} A \Bigg) u_{j} \Bigg],\nonumber
\end{align}
where $f(x)=\sqrt{1+x^2}$ $(x=X,Y,Z)$. The remaining parameters are defined in Table~\ref{tab:2HDM-III_models}.

\begin{widetext}
\centering
\begin{table}[b!]
\begin{tabular}{|c|c|c|c|c|c|c|c|c|c|}
\hline
2HDM-III& $X$ &  $Y$ &  $Z$ & $\xi^u_h $  & $\xi^d_h $ & $\xi^l_{h} $  & $\xi^u_H $  & $\xi^d_H $ & $\xi^{l}_H $\\ \hline
2HDM-I-like~
&  $-\cot\beta$ & $\cot\beta$ & $-\cot\beta$ & $c_\alpha/s_\beta$ & $c_\alpha/s_\beta$ & $c_\alpha/s_\beta$
& $s_\alpha/s_\beta$ & $s_\alpha/s_\beta$ & $s_\alpha/s_\beta$\\
2HDM-II-like
& $\tan\beta$ & $\cot\beta$ & $\tan\beta$ & $c_\alpha/s_\beta$ & $-s_\alpha/c_\beta$ & $-s_\alpha/c_\beta$
& $s_\alpha/s_\beta$ & $c_\alpha/c_\beta$ & $c_\alpha/c_\beta$\\
2HDM-X-like
& $-\cot\beta$ & $\cot\beta$ & $\tan\beta$ &  $c_\alpha/s_\beta$ & $c_\alpha/s_\beta$ & $-s_\alpha/c_\beta$
& $s_\alpha/s_\beta$ & $s_\alpha/s_\beta$ & $c_\alpha/c_\beta$\\
2HDM-Y-like
& $\tan\beta$ & $\cot\beta$ & $-\cot\beta$ & $c_\alpha/s_\beta$ & $-s_\alpha/c_\beta$ & $c_\alpha/s_\beta$
& $s_\alpha/s_\beta$ & $c_\alpha/c_\beta$ & $s_\alpha/s_\beta$\\
\hline
\end{tabular}
\caption{ Parameter expressions for the 2HDM-III like models, where $s_\alpha \equiv \sin\alpha$, $c_\alpha\equiv \cos\alpha$,
$s_\beta\equiv \sin\beta$ y $c_\beta\equiv \cos\beta$.}
\label{tab:2HDM-III_models}
\end{table}
\end{widetext}
The Eq.~\eqref{eq:LYukawa_ns} allows us to write a specific 2HDM plus FC interactions.

To compute the cross-section for $pp\to th+X$, we extract the coupling $g_{tch}$ from Eq.~\eqref{eq:LYukawa_ns}
\begin{equation}
g_{tch}=\frac{g}{2\sqrt{2} m_{W}}
\frac{\left(\xi^{u}_{H}+Y \xi^{u}_{h}\right) \sqrt{m_t m_c}}{f(Y)} \tilde{\chi}_{tc}^{u},
\label{eq:tch_coupling}
\end{equation} 
%
when the $Y$, $\xi^{u}_{H^0}$ and $\xi^{u}_{h^0}$ parameters are replaced in Eq.~\eqref{eq:tch_coupling}, we can see that $g_{tch}$ is the same for all the models of Table~\ref{tab:2HDM-III_models},  meaning that our results are valid for any of these models.

\subsection{Constraints on the 2HDM-III parameter space \label{sec:SecIII}}

We present a general overview of the 2HDM-III parameter space. Details of the analysis can be found in Ref.~\cite{Arroyo-Urena:2024soo}. 

\begin{figure}[!htb]
\centering
\includegraphics[width=12cm]{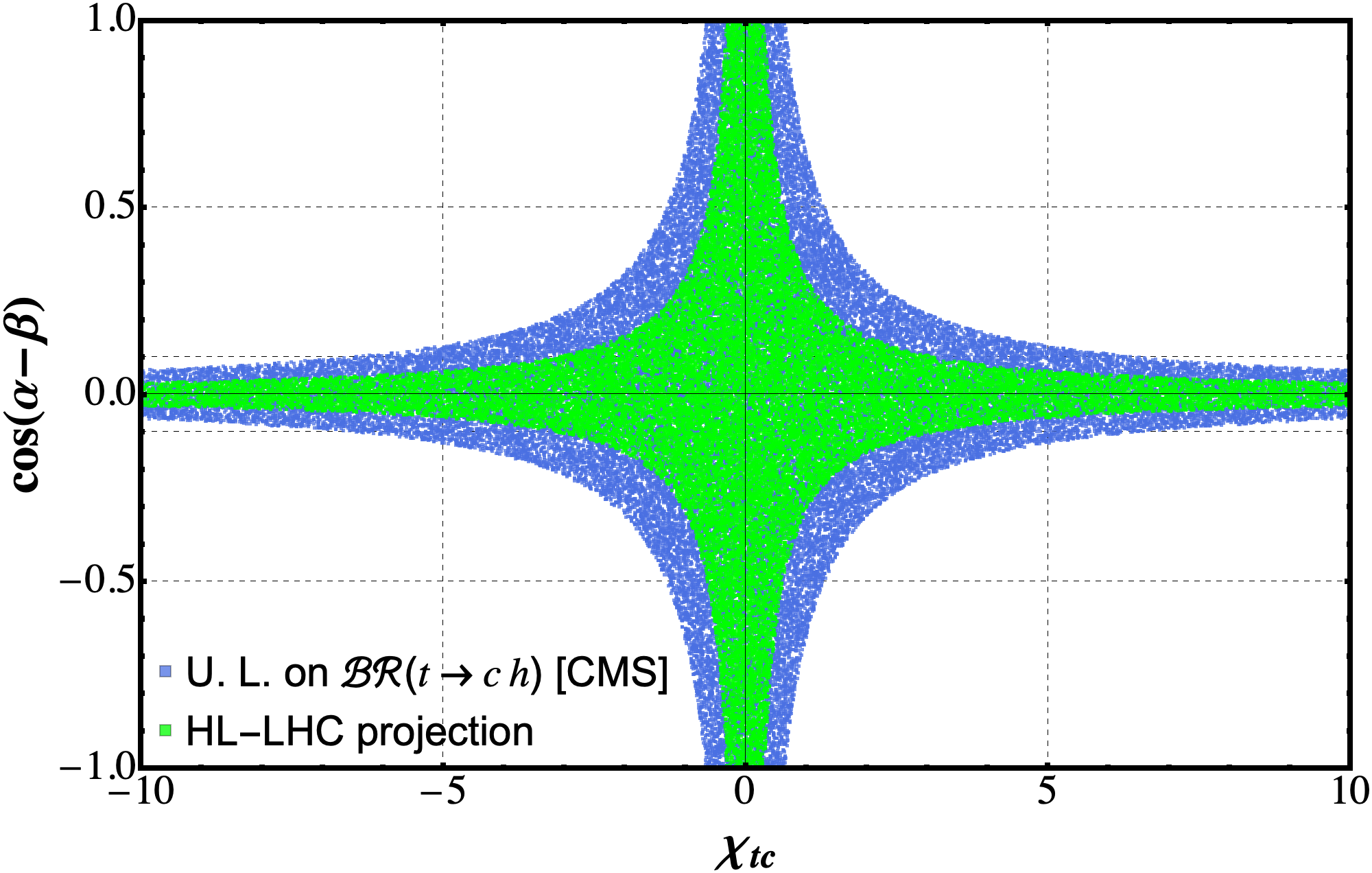}
\caption{Plot of $\chi_{tc}-\cos(\alpha-\beta)$ plane on $\mathcal{BR}(t\to ch)$. The blue points are  allowed by the upper limit given by CMS, while the green points correspond to the HL-LHC projection.}
\label{chi_cosBa} 
\end{figure}

The model free parameters that have a direct impact on our predictions are: 
\begin{enumerate}
\item Cosine of the difference of mixing angles: $\cos(\alpha-\beta)$,
\item Ratio of the VEV's: $\tan\beta$,
\item The parameter that changes flavor $\chi_{tc}$ ($\chi_{tc} \equiv \tilde{\chi}_{tc}^{u}$). 
\end{enumerate}	
The observables we consider to constraint are the following: 
\begin{itemize}
\item LHC Higgs boson data~\cite{CMS:2017con,ATLAS:2019pmk},
\item Neutral meson physics $B_{s}^0\to\mu^+\mu^-$~\cite{CMS:2022mgd}, $B_{d}^0\to\mu^+\mu^-$~\cite{CMS:2022mgd},
\item $\ell_i\to \ell_j\ell_k\bar{\ell}_k$ and $\ell_i\to \ell_j\gamma$~\cite{Workman:2022ynf},
\item Muon anomalous magnetic moment $a_\mu$~\cite{Muong-2:2021ojo}.
\end{itemize}
The only parameter we did not consider in Ref.~\cite{Arroyo-Urena:2024soo} is the matrix element $\chi_{tc}$, which is constrained by the upper limit (U. L.) on $\mathcal{BR}(t\to ch)<0.00043$ reported by CMS~\cite{CMS:2024ubt}, currently more restrictive than the reported by ATLAS~\cite{ATLAS:2022gzn}. We also include the HL-LHC projection on $\mathcal{BR}(t\to ch)<10^{-4}$~\cite{Azzi:2019yne}, at $95\%$ C.L. Fig.~\ref{chi_cosBa} shows the allowed region by CMS and the HL-LHC projection (blue and green points, respectively).

\begin{figure}[!htb]
\centering
\includegraphics[width=12cm]{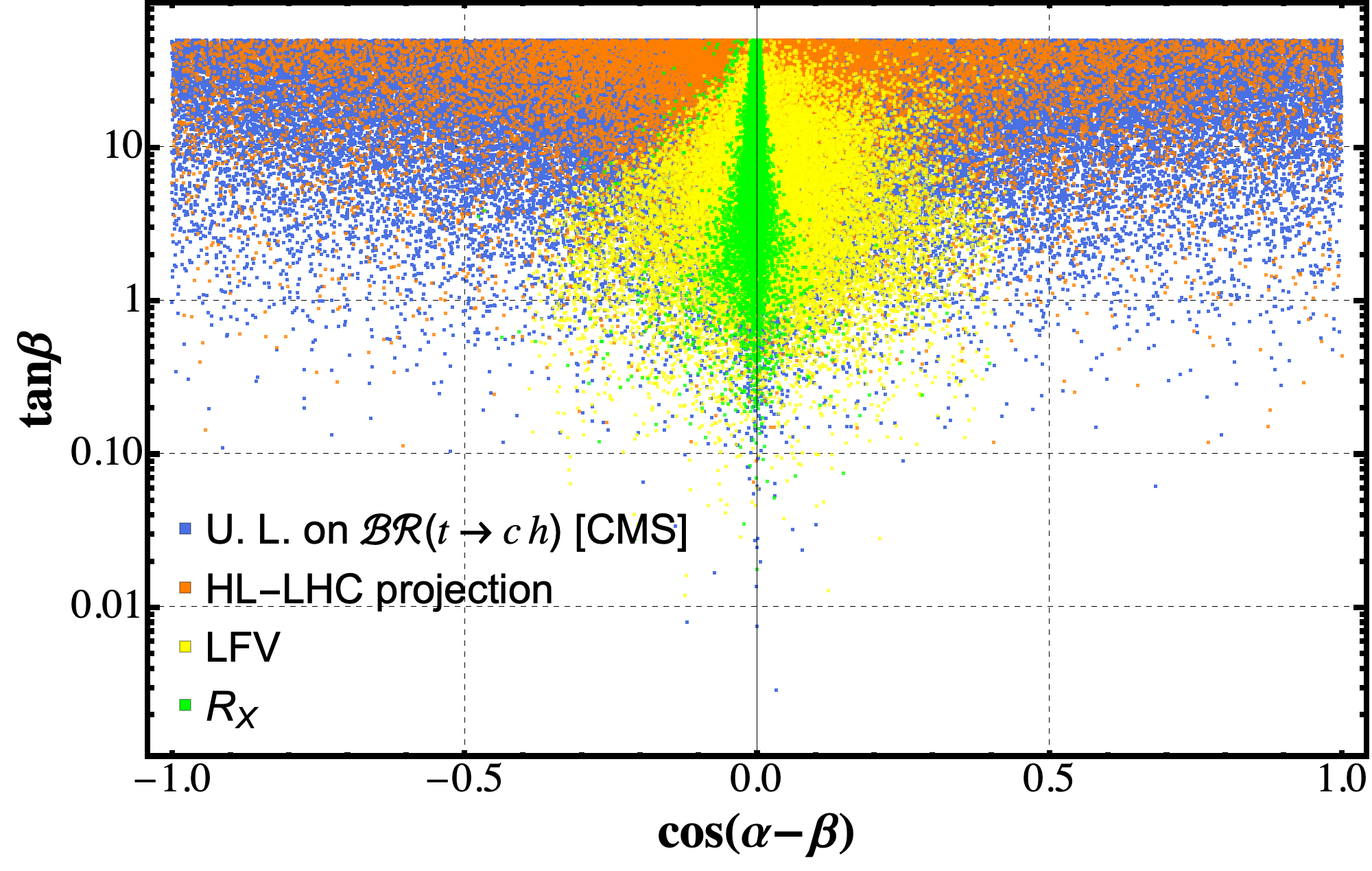}
\caption{Model parameter space of the 2HDM-III in the $\cos(\alpha-\beta)-\tan\beta$ plane. The colored point are those allowed by 1) current upper limit on $\mathcal{BR}(t\to ch)$ (blue points), 2) HL-LHC projection on $\mathcal{BR}(t\to ch)$ (orange points), 3) Lepton Flavor Violating processes (yellow points) and 4) LHC Higgs boson data (green points).}
\label{cosAB_tanB} 
\end{figure}

Meanwhile, in Fig.~\ref{cosAB_tanB} we show the $\cos(\alpha-\beta)-\tan\beta$ plane whose points correspond to those allowed by current upper limit on $\mathcal{BR}(t\to ch)$, HL-LHC projection on $\mathcal{BR}(t\to ch)$, Lepton Flavor Violating processes and LHC Higgs boson data. We observe that the most stringent constraints come from the LHC Higgs boson data, which allow values for $\cos(\alpha-\beta)\sim 0$ corresponding to the interval $0.1<\tan\beta <50$; in agreement with the decoupling limit ($\cos(\alpha-\beta)\to 0$).

The proposed process directly depends on $\chi_{tc}$ and the $Y=\cot\beta$ factor, as shown in Eq.~\eqref{eq:tch_coupling}, it is clear that our signal is favored by small values of $\tan\beta$. 

The upper bound on $\mathcal{BR}(t\to ch)$ has a significant impact on our parameter space. From Fig.~\ref{chi_cosBa}, we observe that the HL-LHC projection reduces the allowed region for the parameters $\cos(\alpha-\beta)$ and $\chi_{tc}$, compared to the one allowed by the CMS U. L. and therefore, lower $\mathcal{BR}(t\to ch)$ values lead to lower cross-sections of the proposed signal.

After analyzing the free model parameter space, we define two scenarios to be used in the next section,
\begin{itemize}
\item $S1:$ $\tan\beta=1$, $\chi_{tc}=1,\,5$ and $\cos(\alpha-\beta)=0.1$,
\item $S2:$ $\tan\beta=3$, $\chi_{tc}=1,\,5$ and $\cos(\alpha-\beta)=0.1$.
\end{itemize}

\section{Collider analysis}\label{sec:SecIV}

In this section, we use a Monte Carlo generator to obtain a sample of simulated events with which we explore the prospects for detecting the proposed signal as well as the SM background processes that obscure it. In order to separate the signal from the background, we use the {\it Boosted Decision Trees} (BDT) machine learning algorithm~\cite{coadou:2022, Hastie:2009itz} instead of making direct kinematic cuts as is commonly done.
\subsection{Signal and background\label{subsec:signal}}
\begin{figure}[!htb]
\centering
\subfigure[]{\includegraphics[width=7cm]{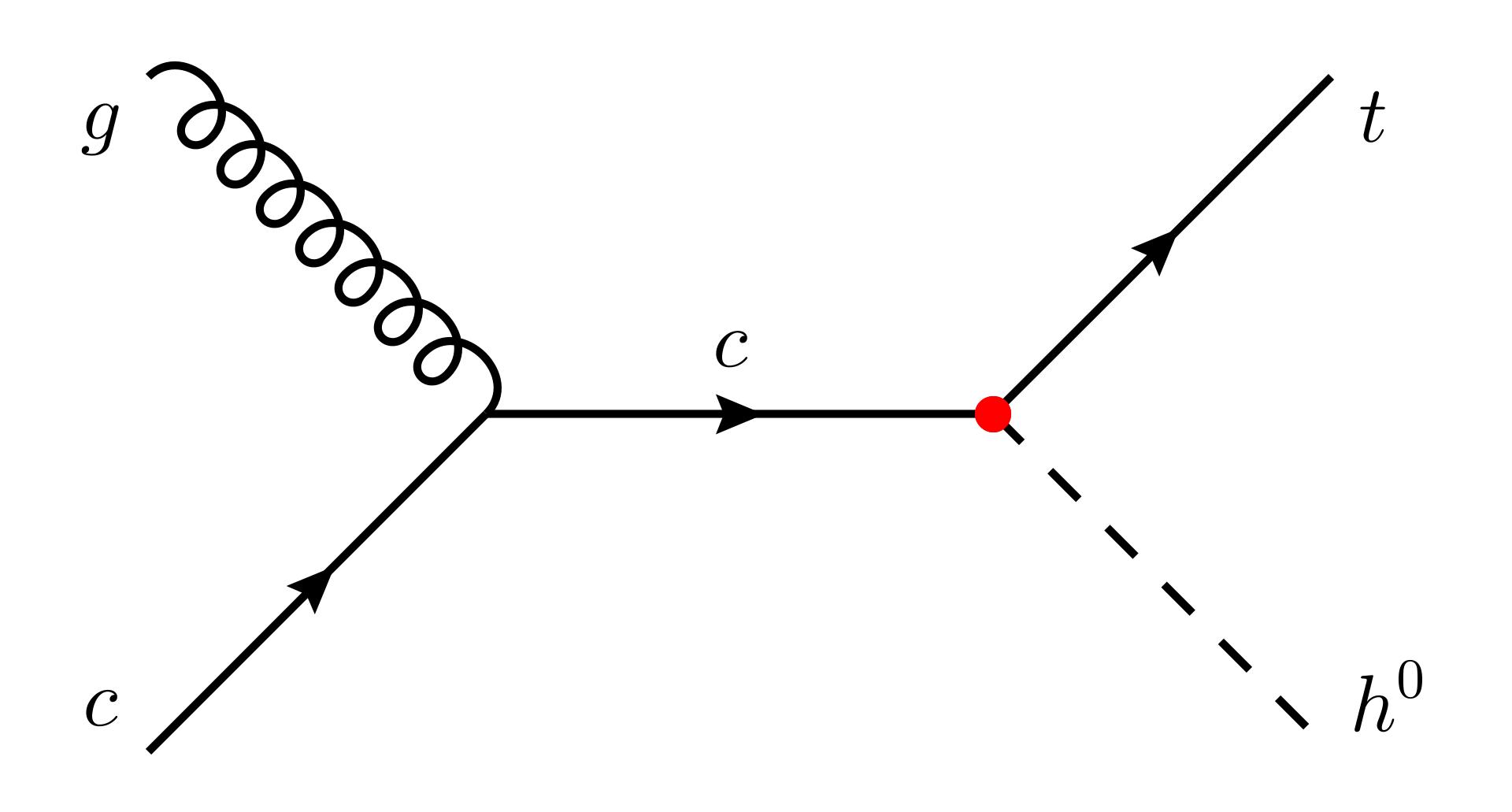}}  
\subfigure[]{\includegraphics[width=5.5cm]{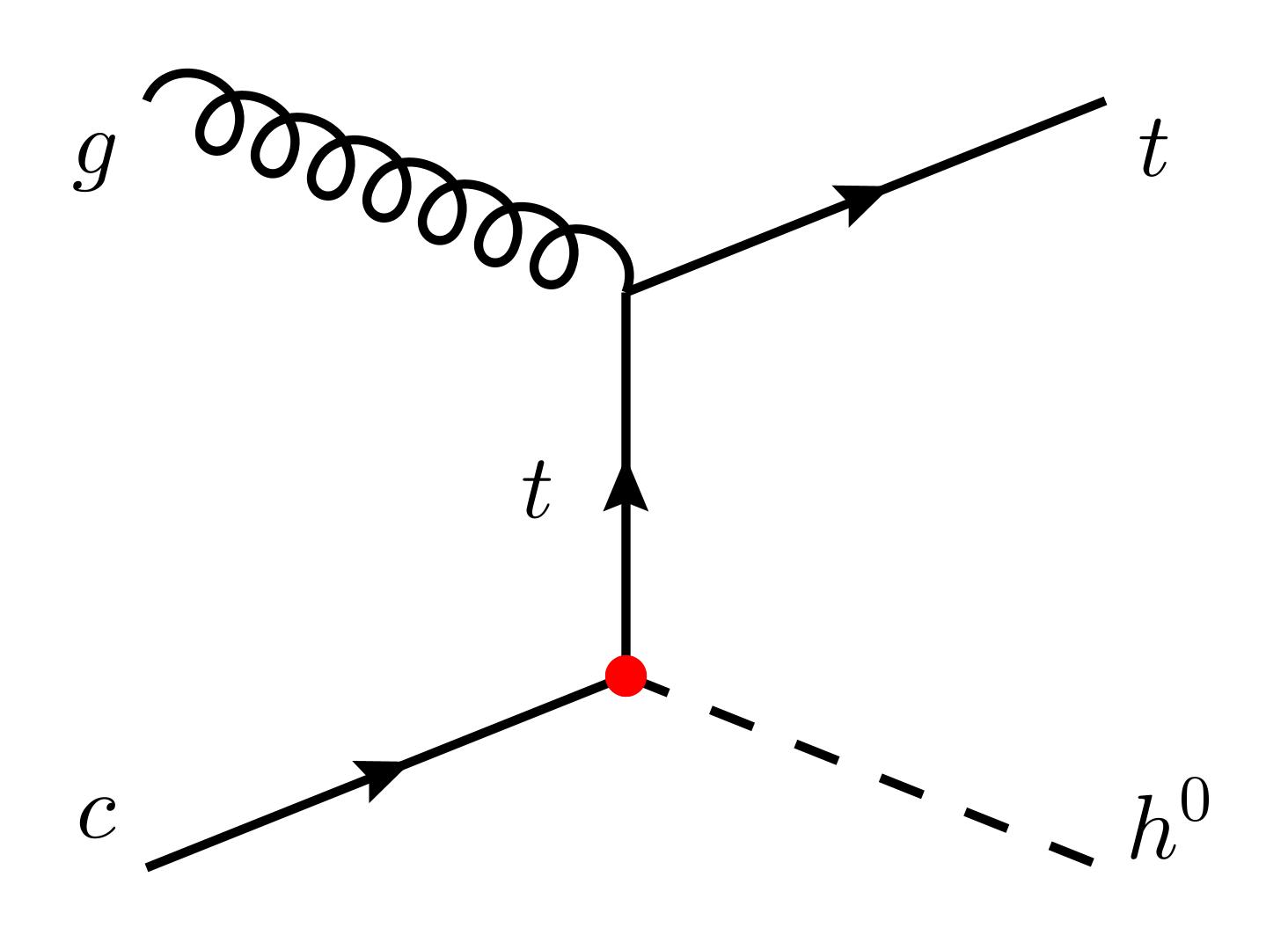}} 
\caption{Feynman diagrams for the process $pp \to th,$ (a) $s$-channel and (b) $t$-channel.}
\label{PPHTdiag}
\end{figure}

We search for the process $pp\to th+X\,(t\to \ell\nu_{\ell}b,\,h\to\gamma\gamma,\,\ell=e,\,\mu)$, whose Feynman diagrams (drawn with \texttt{FeynGame}~\cite{HARLANDER2020107465}) are presented in Fig.~\ref{PPHTdiag}. The red points indicate the new interaction coming from 2HDM-III, which induces the Flavor-Changing interaction. One of the most relevant observables of the signal is the resonant effect that comes from the decay $h\to\gamma\gamma$. Moreover, both photons show a transverse momentum such that their sum is approximately the mass of the Higgs boson, which is of utmost importance for the isolation of the signal. Meanwhile, the main SM background comes from $Whj$, $Wj\gamma\gamma$, $tj\gamma\gamma$, $Wjj\gamma$,
$tj\gamma$,  where $j$ represents non-bottom-jets.

The cross-sections of the signal and the SM background processes are shown in Tables~\ref{XSSignal} and~\ref{XSSMBGD}, respectively.

\begin{table}[!htb]
\caption{Cross-section ($\sigma$) of the signal for the scenarios $S1,\,S2.$}\label{XSSignal}
\begin{centering}
\begin{tabular}{ccc}
\hline 
Scenario & $\sigma$ {[}fb{]} $(\chi_{tc}=1)$ & $\sigma$ {[}fb{]} $(\chi_{tc}=5)$ \tabularnewline
\hline 
\hline 
$S1$ & 0.01 & 0.025 \tabularnewline
\hline 
$S2$ & 0.004 & 0.014 \tabularnewline
\hline 
\end{tabular}
\par\end{centering}
\end{table}
\begin{table}[!htb]
\caption{Cross-section of the dominant SM background processes.}\label{XSSMBGD}
\begin{centering}
\begin{tabular}{cc}
\hline 
SM backgrounds & Cross-section {[}fb{]}\tabularnewline
\hline 
\hline 
$pp\to Whj$ & $0.19$\tabularnewline
\hline 
$pp\to Wj\gamma\gamma$ & $68.27$\tabularnewline
\hline 
$pp\to tj\gamma\gamma$ & $0.019$\tabularnewline
\hline 
$pp\to Wjj\gamma$ & $16590$\tabularnewline
\hline 
$pp\to tj\gamma$ & $1.84$\tabularnewline
\hline 
\end{tabular}
\par\end{centering}
\end{table}

As far as our computation scheme is concerned, we first implement the model via \texttt{FeynRules}~\cite{Alloul:2013bka} for the \texttt{MadGraph5}~\cite{Alwall:2014hca} event generator, interfaced with \texttt{Pythia8}~\cite{Sjostrand:2014zea} for parton showering and \texttt{Delphes3}~\cite{deFavereau:2013fsa} for detector simulations (using the HL-LHC card~\cite{HL-LHC_card}). Concerning to the jet reconstruction, the finding package \texttt{FastJet}~\cite{Cacciari:2011ma} and the \texttt{anti-$k_t$} algorithm~\cite{Cacciari:2008gp} package were used. We also include the $b$-tagging efficiency $\epsilon_b=90\%$. The probability that a $c-$jet or any other light-jet $j$ is mistagged as a $b-$jet are $\epsilon_c=5\%$~\cite{ATLAS:2023gog} and $\epsilon_j=1\%$, respectively.

\begin{figure}[!htb]
\begin{center}
\subfigure[]{\includegraphics[scale=0.45]{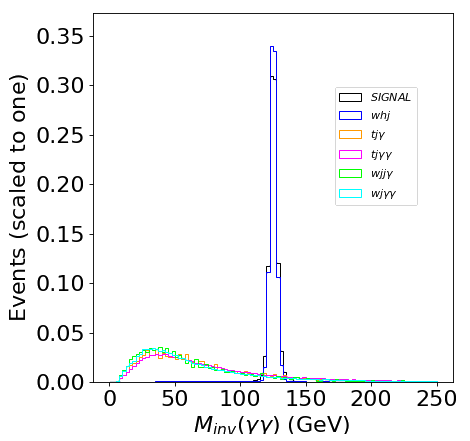}}
\subfigure[]{\includegraphics[scale=0.45]{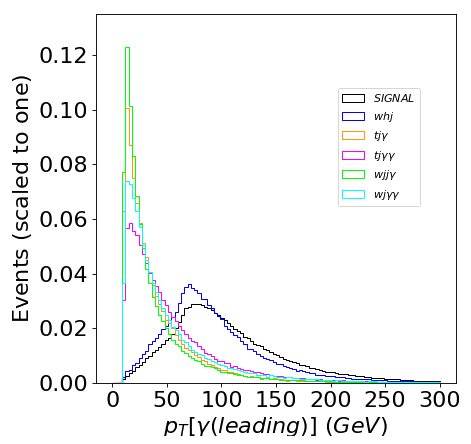}}
\subfigure[]{\includegraphics[scale=0.45]{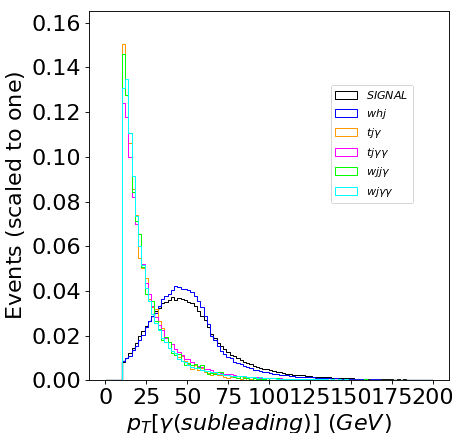}}
\end{center}
\caption{Plots for the signal and background variables: (a) Invariant mass of the two photons, (b) $p_T$ of the leading photon, and (c) $p_T$ of the subleading photon.}
\label{distributions}
\end{figure}

The traditional strategy of making hard cuts on the observables removes a significant number of background events but also rejects an important number of signal events. This situation can be improved by using MVA techniques, such as BDT. We perform a BDT training using variables related to the kinematics of the final state, including the transverse momentum $(p_T)$ and the pseudo-rapidities of the $b$-jet, the charged lepton $\ell$ and photons. We present in Fig.~\ref{distributions} the most discriminating training variables, plotted with \texttt{MadAnalysis5}~\cite{Conte:2012fm}.
The relevant hyperparameters for the BDT training are as follows: Number of trees \texttt{NTree}=50, maximum depth of the decision tree \texttt{MaxDepth}=5, maximum number of leaves \texttt{MaxLeaves}=8; the remaining parameters are set to their default values.

\begin{figure}[!htb]
\centering
\includegraphics[scale=0.5]{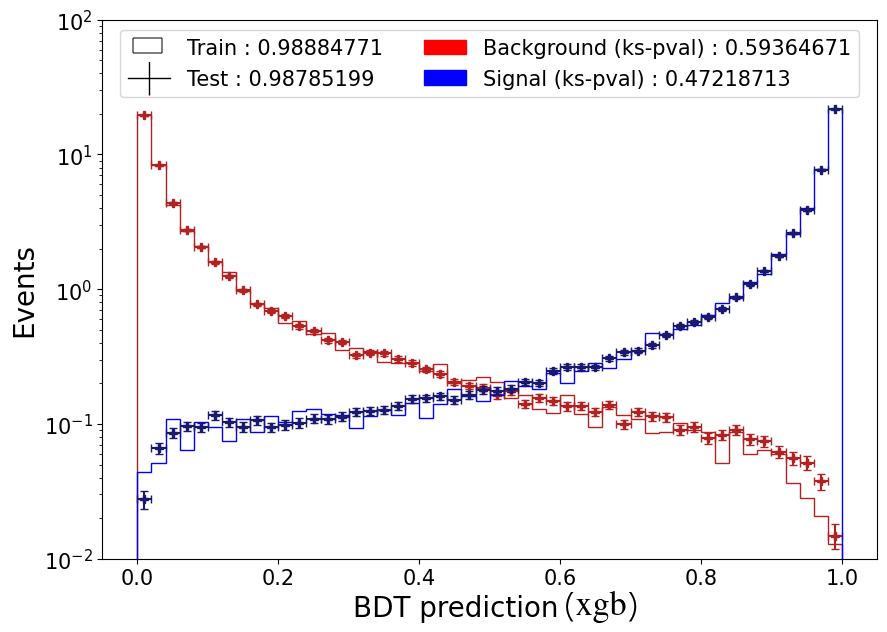}
\caption{Plot of the discriminant for signal and background data.}
\label{Discriminat}
\end{figure}

We present in Fig.~\ref{Discriminat} the discriminant for the signal and background. The goodness of fit is verified with the Kolmogorov-Smirnov (KS) test. We observed that the KS value lies within the allowed interval $[0,1]$ and has a value of 0.47 (0.59) for the signal (background). We therefore conclude the training did not exhibit signs of overtraining.
Once the classifier has been trained, it exports the output in terms of a single variable $(\textbf{xgb})$, which separates the signal events from the background, as shown in Fig.~\ref{Discriminat}. Then, the signal-to-background ratio is optimized. The BDT training is computed using the Monte Carlo simulated data. The signal and background samples are scaled to the expected number of events, which
is calculated via the integrated luminosity and cross-section values. The BDT selection is optimized individually for each channel to maximize the figure of merit, {\it i.e.}, the \textit{signal significance}, defined as $S/\sqrt{S + B+(\kappa\cdot B)^2}$~\cite{Cowan2021}, where $S$ and $B$ represent the number of signal and background candidates, respectively. The factor $\kappa$ represents a realistic $5\%$ systematic uncertainty in the SM background estimation~\cite{ATL-PHYS-PUB-2016-019}.

\begin{figure}[!htb]
\centering
\subfigure[]{\includegraphics[scale=0.27]{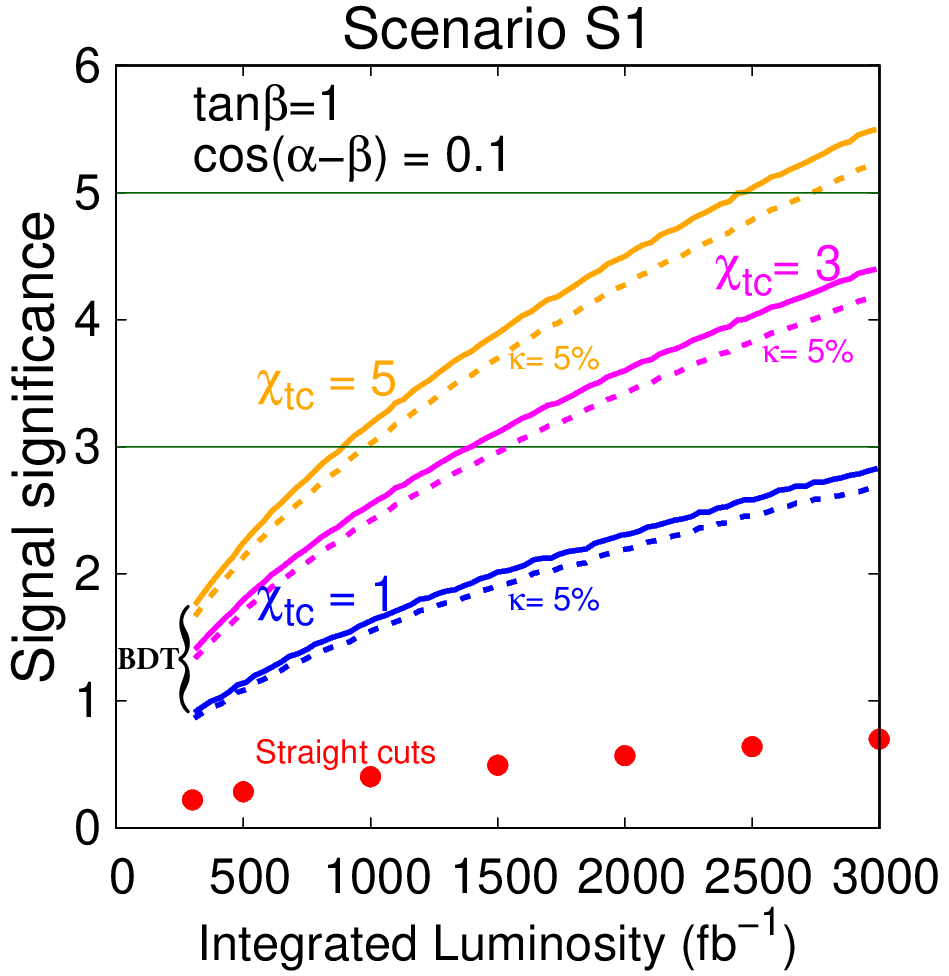}}
\subfigure[]{\includegraphics[scale=0.27]{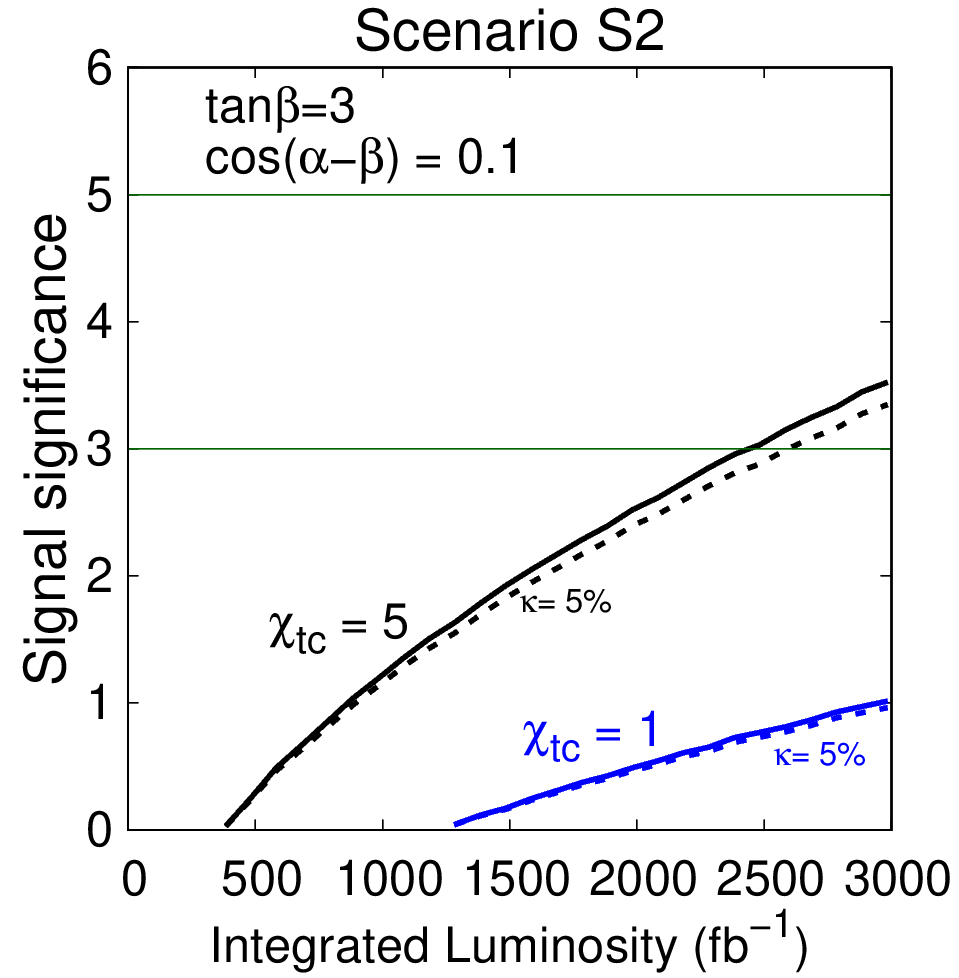}}
\caption{\textit{Signal significance} as a function of the integrated luminosity: (a) Scenario $S1$, (b) Scenario $S2$. We have considered a systematic uncertainty  $\kappa=5\%$ (dashed lines), solid lines do not consider it. The red dots in (a) correspond to the \textit{signal significance} for $\chi_{tc}=5$, calculated through "straight" cuts on basic variables, as described in the main text.} 
\label{signalsignificances}
\end{figure}

Fig.~\ref{signalsignificances} presents our main results, \textit{i.e}, the \textit{signal significance} as a function of the integrated luminosity for the two scenarios defined in the previous section: $S1,\,S2$. To compare the results obtained via BDT versus the obtained through \texttt{MadAnalysis5}, we also include in Fig.~\ref{signalsignificances} (a) the plot for the \textit{signal significance} imposing ``straight" kinematic cuts on the variables (red dots), using the following cuts:
\begin{itemize}
\item Invariant mass of the two photons $120<M_{\rm inv}(\gamma\gamma)<130$ GeV,
\item Transverse mass $120<M_T<190$ GeV, where $M_T^2=\Big(\sqrt{(p_\ell+p_b)^2+|\vec{p}_{T,\ell}+\vec{p}_{T,b}|^2}+|\vec{\textbf{P}}_T|\Big)^2-|\vec{p}_{T,\ell}+\vec{p}_{T,b}+\vec{\textbf{P}}_T|^2$, with $\vec{\textbf{P}}_T$ being the missing transverse momentum given by the negative sum of visible momenta in the transverse direction.
\item $P_T[\gamma(\rm leading)]>50$ GeV and $P_T[\gamma(\rm subleading)]>30$ GeV.
\end{itemize}
It is clear that the use of BDT in our analysis substantially increases the \textit{signal significance}. 

We find that the scenario $S1$ is the most promising, this because the cross-section for $pp\to th+X$ is highly sensitive to low values of $\tan\beta$ ($g_{htc}\sim 1/\tan\beta$), as shown in Eq.~\eqref{eq:LYukawa_ns}.  In this case, we predict a \textit{signal significance} $\sigma\geq 5$ ($\sigma= 3$) for an integrated luminosity $\mathcal{L}_{\rm int}\geq 2500$ fb$^{-1}$ ($\mathcal{L}_{\rm int}\sim 900$ fb$^{-1}$), and $\chi_{tc}=5$. The situation changes slightly when including systematic uncertainties $\kappa=5\%$: $\sigma\geq 5$ ($\sigma= 3$) for $\mathcal{L}_{\rm int}\geq 2700$ fb$^{-1}$ ($\mathcal{L}_{\rm int}\sim 1000$ fb$^{-1}$). To satisfy the allowed region by the HL-LHC projection on the upper limit of $\mathcal{BR}(t\to ch)$, we add $\chi_{tc}=3$ to the scenario $S1$. Thus, we obtain a \textit{signal significance} $\sigma\approx 4.4 \,(4.2)$ for $\kappa=0$ ($\kappa=5\%$) and $\mathcal{L}_{\rm int}= 3000$ fb$^{-1}$. These results indicate that in the  HL-LHC experiment, evidence for the process $pp\to th+X$ could be achieved. Meanwhile, scenario $S2$ is the less favored because we set $\tan\beta=3$; although the \textit{signal significance} for $S2$ is relatively low, evidence for $pp\to th+X$ can be achieved if a $\mathcal{L}_{\rm int}\sim 2500$ fb$^{-1}$ is reached.


\section{Summary and conclusions}\label{sec:SecV}

The search for new physics is fundamental in collider experiments such as the LHC, particularly interesting are the Flavor-Violating processes which are very suppressed in the SM. The 2HDM-III allows for this kind of interactions at tree-level, which significantly increasing their production cross-sections. In this work, we have investigated the Flavor-Violating process $pp \rightarrow th+X$, which does not exist in the SM but has origin in the $htc$ interaction predicted in the theoretical framework of the 2HDM-III. 

We first have presented the relevant aspects of the 2HDM-III in which we studied prospects for detecting a single Higgs boson in association with a top quark. Specifically, we search for the final state $t\to \ell\nu_\ell b$ and $h\to\gamma\gamma$, whose kinematic characteristics of the photons provided valuable information to isolate the signal from the background. The model parameter space was constrained on the basis of experimental results, namely, LHC Higgs boson data, upper limits on $\mathcal{BR}(t\rightarrow ch)$ and relevant lepton FV processes. Using the previous results, we define two realistic scenarios $S1,\,S2$. In order to increase the \textit{signal significance}, we performed a Multivariate Analysis using BDT for the two scenarios. After using BDT, we found that the scenario $S1$: $\tan\beta=1$, $\chi_{tc}=1,\,3,\,5$ and $\cos(\alpha-\beta)=0.1$, is the most promising, which is expected since the $htc\sim 1/\tan\beta$ interaction is favored for small values of $\tan\beta$. Our results allow us to predict a \textit{signal significance} at level of $\geq 5\sigma$ once the integrated luminosity $\mathcal{L}_{\rm int}\gtrsim2500~fb^{-1}$ is reached. If we consider a systematic uncertainty of $\kappa=5\%$, $\mathcal{L}_{\rm int}\gtrsim2700~fb^{-1}$ would be necessary. Taking into account the HL-LHC projection on $\mathcal{BR}(t\to ch)$, we obtain a \textit{signal significance} $\sigma \approx 4.4$ for an integrated luminosity $\mathcal{L}_{\rm int}=3000~fb^{-1}$. This result decreases slightly to $\sigma\approx4.2$ if a systematic uncertainty of $\kappa=5\%$ is considered. Given our results for the proposed signal, it is very likely that a future collider with higher center-of-mass energy may permit to achieve a higher significance for a broad range of parameters.

\section*{Acknowledgements}
The work of Marco A. Arroyo-Ure\~na and T. Valencia-P\'erez is supported by Estancias Posdoctorales por M\'exico SECIHTI (formerly CONAHCYT) and Sistema Nacional de Investigadores e Investigadoras (SNII-SECIHTI). T.V.P. acknowledges support from the UNAM project 
PAPIIT No IN111224 and the SECIHTI project CBF2023-2024-548. V.M. L\'opez-Guerrero thanks to SECIHTI M\'exico for the PhD fellowship. J.L. D\'iaz-Cruz and O. F\'elix-Beltr\'an. thank the support of SNII-SECIHTI and Vicerrector\'ia de Investigaci\'on y Estudios de Posgrado (VIEP-BUAP).

\appendix
 


\end{document}